\RequirePackage{rotating}
\documentclass[a4paper,fleqn,usenatbib]{mnras}

\usepackage{mathptmx}
\usepackage{morefloats}

\usepackage[T1]{fontenc}
\usepackage{ae,aecompl}

\usepackage{graphicx}	
\usepackage{amsmath}	
\usepackage{amssymb}	
\usepackage{multicol}

\usepackage{color}
\usepackage{subfig}
\usepackage{float}
\usepackage{rotating}

\title[The warm DQ OW J175358.85-310728.9]{The OmegaWhite Survey for Short-Period Variable Stars IV: Discovery of the warm DQ white dwarf OW$\,$J175358.85--310728.9}

\author[S. Macfarlane et al.]  
{S.A~Macfarlane$^{1,2}$,
  P.A~Woudt$^2$\thanks{Email: pwoudt@ast.uct.ac.za},
  P.~Dufour$^3$,  
  G.~Ramsay$^4$,
  P.J~Groot$^1$,
  R.~Toma$^4$,
  \newauthor
  B.~Warner$^2$,
  K.~Paterson$^2$,
  T.~Kupfer$^5$, 
  J.~van Roestel$^1$, 
  L.~Berdnikov$^{6,7,8}$,
  \newauthor
  T.~Dagne$^6$,
  F.~Hardy$^3$
  \\
  $^1$Department of Astrophysics/IMAPP, 
      Radboud University, 
      P.O. Box 9010,
      6500 GL Nijmegen,
      The Netherlands \\
  $^2$Department of Astronomy, 
      University of Cape Town, 
      Private Bag X3,
      Rondebosch 7701, 
      South Africa \\   
  $^3$ Departement de Physique, Universite de Montreal,
      C.P. 6128,
      Succ. Centre-Ville,
      Montreal,
      QC H3C 3J7,
      Canada   \\    
  $^4$Armagh Observatory, 
      College Hill, 
      Armagh, 
      BT61 9DG,
      Northern Ireland, 
      UK\\
  $^5$Division of Physics, Mathematics, and Astronomy
      California Institute of Technology
      Pasadena
      CA 91125
      USA \\         
  $^6$Astronomy and Astrophysics Research division, 
      Entoto Observatory and Research Center, 
      PO Box 8412 Addis Ababa, 
      Ethiopia\\
  $^7$Sternberg Astronomical Institute, 
      Lomonosov Moscow State University, 
      Universitetskij Pr. 13, 
      Moscow 119992, 
      Russia\\
  $^8$Isaac Newton Institute of Chile, 
      Moscow Branch, 
      Universitetskij Pr. 13, 
      Moscow 119992, 
      Russia\\                        
}

\date{Accepted XXX. Received YYY; in original form ZZZ}

\pubyear{2017}

\begin{document}
\label{firstpage}
\pagerange{\pageref{firstpage}--\pageref{lastpage}}
\maketitle

\begin{abstract}
  We present the discovery and follow-up observations of the second known \textcolor{black}{variable} warm DQ white dwarf OW$\,$J175358.85--310728.9 (OW J1753--3107). OW J1753--3107 is the brightest of any of the currently known warm or hot DQ and was discovered in the OmegaWhite Survey as exhibiting optical variations on a period of 35.5452 (2) mins, with no evidence for other periods in its light curves. This period has remained constant over the last two years and a single-period sinusoidal model provides a good fit for all follow-up light curves. The spectrum consists of a very blue continuum with strong absorption lines of neutral and ionised carbon, a broad He\,I $\lambda$4471 line, and possibly weaker hydrogen lines. The C\,I lines are Zeeman split, and indicate the presence of a strong magnetic field. Using spectral Paschen-Back model descriptions, we determine that OW J1753--3107 exhibits the following physical parameters: $T_{\rm eff} =$ \textcolor{black}{15430} K, $\log{(g)} =$ \textcolor{black}{9.0}, $\log{(N(C)/N(He))} =$ \textcolor{black}{--1.2}, and the mean magnetic field strength is B$_{z} =$2.1 MG. This relatively low temperature and carbon abundance (compared to the expected properties of hot DQs) is similar to that seen in the other warm DQ SDSS$\,$J1036$+$6522. Although OW J1753--3107 appears to be a twin of SDSS$\,$J1036$+$6522, it exhibits a modulation on a period slightly longer than the dominant period in  SDSS$\,$J1036$+$6522 and has a higher carbon abundance. The source of variations is uncertain, but they are believed to originate from the rotation of the magnetic white dwarf.
  
\end{abstract}




\section{Introduction}

White dwarfs (WD) have been shown to exhibit short-period variations since the mid-20th century, when the first variable white dwarf, HL Tau 76, was observed to pulsate on a period of 12.5 minutes \citep{Landolt1968}.  HL Tau 76 is now known as a white dwarf variable containing a hydrogen-enriched atmosphere (DAV, or  ZZ Ceti stars) and has since been observed to pulsate in multiple modes \citep[e.g.][]{Dolez2006}. Other variable types of white dwarf stars include helium-dominated WDs (DBV, or V777 Her), WDs with atmospheres that are rich in carbon, helium, and oxygen \citep[GW Vir stars, see ][for detailed reviews of pulsating WD]{Winget2008,Althaus2010,Fontaine2008}, extremely low mass white dwarfs \citep[ELMV,][]{Hermes2012} and a new class of white dwarfs that exhibit carbon-enriched atmospheres (``hot DQ", \citealt{Montgomery2008} or ``warm DQ", \citealt{Williams2013}).

Although the variations exhibited in DAV and DBV stars can be attributed to non-radial gravity wave pulsations, it has recently been shown that the origin of variations in warm and hot carbon-enriched DQs are likely due to the rotation of the white dwarf \citep[for e.g.][]{Williams2016}. As only eight of these warm and hot DQ WD have so far been observed to show periodic variations, it is important that more objects are found and studied. Such research can then be used to test the proposed evolutionary channels of (DQ) white dwarfs. 

The typical cool DQ white dwarf exhibits neutral carbon or molecular C$_{2}$ (Swan bands) in its spectra, and a temperature below $\sim$13000 K \citep[see for example,][]{Dufour2005}. However, \citet{Liebert2003} discovered a new class of DQ in the Sloan Digital Sky Survey \citep[SDSS,][]{Alam2015} that exhibit neutral and/or ionized atomic Carbon lines in their spectra. Using this sample, as well as another similarly hot DQ discovered by \citet{Williams2006}, \citet{Dufour2007,Dufour2008a} found that these so called ``hot DQ'' WDs have carbon-dominated atmospheres, a carbon abundance $\log{(N(C)/N(He))} \geq$ 1, and all lie within a specific temperature range (18000\,K $<$ $T_{\rm eff}$ $<$ 24000 K). Furthermore, $\sim$70$\%$ of these hot DQs have been found to be strongly magnetic with magnetic field strengths $\geq$1 MG \citep{Dufour2011,Williams2016}.

Following the discovery of the first six hot DQs \citep{Montgomery2008,Dunlap2010,Barlow2008,Dufour2011,Lawrie2013}, a seventh DQ, namely SDSS$\,$J103655.39$+$652252.2  \citep[hereafter SDSS$\,$J1036$+$6522,][]{Williams2013}, was discovered at a cooler temperature and with a lower carbon abundance than that of the previously discovered hot DQs ($T_{\rm eff}$ $\approx$ 15500 K, $\log{(N(C)/N(He))} =-1.0$). Thus, it is believed to be a ``warm" DQ, in a transition state from the hot DQs to a set of cooler helium-dominated DQ WD with a temperature in the range 13000\,K $<$ $T_{\rm eff}$ $<$ 18000 K \citep{Dufour2013,Williams2013,Fortier2015}.

In this paper, we present the discovery and initial follow-up observations of a second \textcolor{black}{variable} warm DQ, OW$\,$J175358.85-310728.9 (hereafter OW$\,$J1753--3107) discovered in the OmegaWhite Survey \citep{Macfarlane2015}. As this is the \textcolor{black}{second} warm or hot DQ to be discovered outside the SDSS survey \citep[see also,][]{Williams2006}, in Section~\ref{SEC:Discovery} we discuss details of the discovery and describe the survey in which it was found. We present photometric follow-up observations in Section~\ref{SEC:Photometry}, where we analyse the light curves for periods and compare pulse shape properties spanning 2 years of observations. In Section~\ref{SEC:Spectroscopy}, spectral lines are identified in a high-resolution spectrum of OW$\,$J1753--3107, and results from the modelling of spectral lines are presented. Lastly, we discuss the (magnetic) nature of \textcolor{black}{OW$\,$J1753--3107}, as well as the origin of variations.

\section{Discovery of DQ OW J1753--3107}
\label{SEC:Discovery}

\begin{figure*}
\centering
\includegraphics[width=\linewidth]{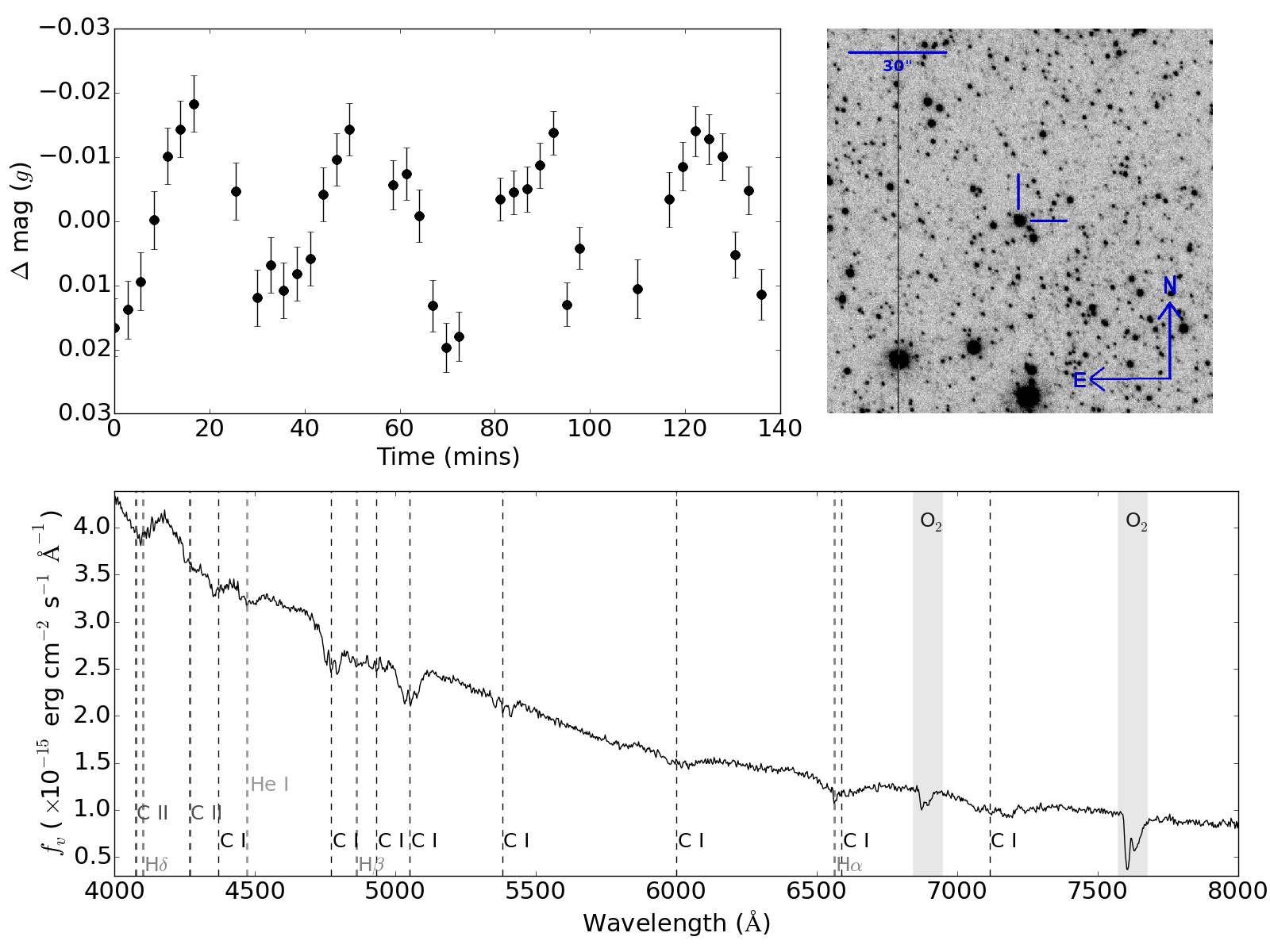}
\caption{$g$-band OmegaWhite light curve (upper left panel), finding chart (upper right panel), and SpUpNIC-observed spectrum (lower panel) of the warm DQ OW$\,$J1753--3107. The finding chart is a 2\arcmin$\times$2\arcmin\ subframe of an OW reference frame, observed in the $g$-band. The spectrum has been flux and wavelength calibrated. Dashed lines indicate important spectral lines (as labelled), and shaded areas cover atmospheric telluric bands.}
\label{FIG_discovery}
\end{figure*}

\begin{table}
\centering
\caption{Target Properties for OW$\,$J1753--3107: star ID; RA and Dec; OW Lomb Scargle period (P$_{OW}$), OW calibrated $g$-band magnitude (OW$_{g}$), amplitude of OW light curve ($A_{OW}$), VPHAS+ colour indices $g - r$ and $u - g$. All instrumental magnitudes in OW are calibrated using the AAVSO All-Sky Survey \citep{Henden2012} to the standard (Vega) scale.}
\begin{tabular}[pos]{ll}  
\hline
\bf Star ID & \bf OW$\,$J175358.85--310728.9 \\ 
\hline
RA (J2000) &  17h53m58.85s \\ 
DEC (J2000) & --31\degr07\arcmin28.9\arcsec \\
OW$_{g}$ & 15.8 mag \\
P$_{OW}$ &  35.2 mins \\
$A_{OW}$ & 2.6 $\%$\\
\hline
\multicolumn{2}{l}{VPHAS+ Colours, on standard (Vega) scale}\\
\hline
$u$ & 14.399 $\pm$ 0.003\\
$g$ & 15.735 $\pm$ 0.005\\
$r^{*}$ & 15.890 $\pm$  0.008\\
$i$ & 15.796 $\pm$  0.008\\
$g-r$ & --0.155 $\pm$ 0.013 \\
$u-g$ & --1.336 $\pm$ 0.008 \\
\hline
\multicolumn{2}{p{6cm}}{$^{*}$mean r-band colour obtained from two separate VPHAS+ epochs \citep{Drew2014}}
\end{tabular}
\label{TAB:properties}
\end{table}

The OmegaWhite (OW) Survey \citep{Macfarlane2015,Toma2016} is a wide-field high-cadence survey which aims to cover 400 square degrees of the Galactic Plane with $g$-band observations, reaching a depth of 21.5 mag. OW sources are cross-matched with the VST Photometric H$\alpha$ Survey of the Southern Galactic Plane \citep[VPHAS+,][]{Drew2014} in order to obtain multi-band colour information. Furthermore, all instrumental magnitudes in OW are calibrated to standard Vega magnitudes using the AAVSO All-Sky Survey \citep{Henden2012}. The aim of the OW survey is to identify short-period variable systems, varying on timescales of minutes to less than an hour, such as ultracompact binary star systems (UCB) and rare pulsating white dwarf/subdwarf sources. Each square degree field is observed with 39-second exposures over a 2-hr duration \citep[\textcolor{black}{with a mean cadence of $3.6 \pm 0.7$ mins,}][]{Toma2016} using OmegaCAM on the VLT Survey Telescope \citep[VST,][]{Capaccioli2011,Kuijken2002} 

OW$\,$J1753--3107 was initially selected for follow-up observations as it exhibited very blue colours ($g-r =$ --0.16, $u-g =$ --1.33) and short-period variations (P$_{OW} \approx $ 35 mins) typically associated with AM CVn systems (see Table~\ref{TAB:properties} for target properties).  In Figure~\ref{FIG_discovery}, we show the OW light curve, along with a finding chart created using a subframe of the OW $g$-band reference frame containing OW$\,$J1753--3107. 

Using the Spectrograph Upgrade -- Newly Improved Cassegrain spectrograph (SpUpNIC, Crause et al. in prep) on the South African Astronomical Observatory (SAAO) 1.9-m telescope on 30 April 2016, we obtained two 1800-s longslit exposures using a slit width of 1.5$\arcsec$ with grating 7 for a dispersion of 2.8 \AA/pixel across the wavelength range ($\sim$3500 \AA\ -- 8000 \AA). The median-combined, flux-calibrated spectrum, shown in Figure~\ref{FIG_discovery}, reveals a blue continuum, along with evidence for absorption lines of C\,I, C\,II, He\,I $\lambda$4471, and possibly H$\alpha$. Furthermore, the presence of Zeeman split lines is evidence that OW$\,$J1753--3107 contains a relatively strong magnetic field. Further photometric and spectroscopic observations of OW$\,$J1753--3107 were carried out in order to precisely characterise the nature of this DQ WD source.

\section{Photometric Analysis}
\label{SEC:Photometry}

\begin{table*}
\centering
\caption{Photometric observing log with parameters: run identification name (Run Id), date of observation (Date Obs.), time at the start of the observation (HJD$_{0}$), the telescope (Tel.), instrument (Instr.) and filter used, the number of pixels binned in each image (binning), the length of each exposure (Exp. Time), and the duration of the observing run (duration) }
\label{TAB_PhotObsLog}
\begin{tabular}{lllllllll}
\hline
Run ID 			& Date Obs. 		& HJD$_{0}$             & Tel.  		& Instr. 	& Filter      & Binning & Exp. Time & Duration \\
				& (dd/mm/yyyy)  &              &            &            &             &         & (s)       & (hrs)    \\
\hline
LC$\_$OW 		& 29/07/2014    & 2456867.9920   & VST        & OmegaCAM   & $g^{1}$    & 1x1     & 39$^{2}$  & 2.3      \\
LC$\_$SHOC$\_$1 & 21/06/2015    & 2457195.3333 & SAAO 1.9-m & SHOC       & white light & 6x6     & 5         & 3.0        \\
LC$\_$SHOC$\_$2 & 16/05/2016     & 2457525.4137 & SAAO 1.0-m & SHOC       & white light & 1x1     & 10, 15    & 6.4      \\
LC$\_$SHOC$\_$3 & 08/06/2016      & 2457548.3247 & SAAO 1.0-m & SHOC       & B           & 2x2     & 30        & 7.9      \\
LC$\_$SHOC$\_$4 & 12/06/2016     & 2457552.3227 & SAAO 1.0-m & SHOC       & B           & 2x2     & 60, 30    & 6.8      \\
LC$\_$SHOC$\_$5 & 13/07/2016      & 2457583.4301 & SAAO 1.0-m & SHOC       & B           & 2x2     & 30        & 4.2      \\
LC$\_$SHOC$\_$6 & 14/07/2016     & 2457584.2310 & SAAO 1.0-m & SHOC       & B           & 2x2     & 30    & 8.0      \\
LC$\_$SHOC$\_$7 & 16/07/2016      & 2457586.3603 & SAAO 1.0-m & SHOC       & B           & 2x2     & 45        & 4.8      \\
\hline
\multicolumn{9}{p{17.0cm}}{$^{1}$OW magnitude in standard (Vega) system, $^{2}$with a cadence of $\sim$167 s}
\end{tabular}
\end{table*}

\begin{figure}
\centering
\includegraphics[width=\linewidth]{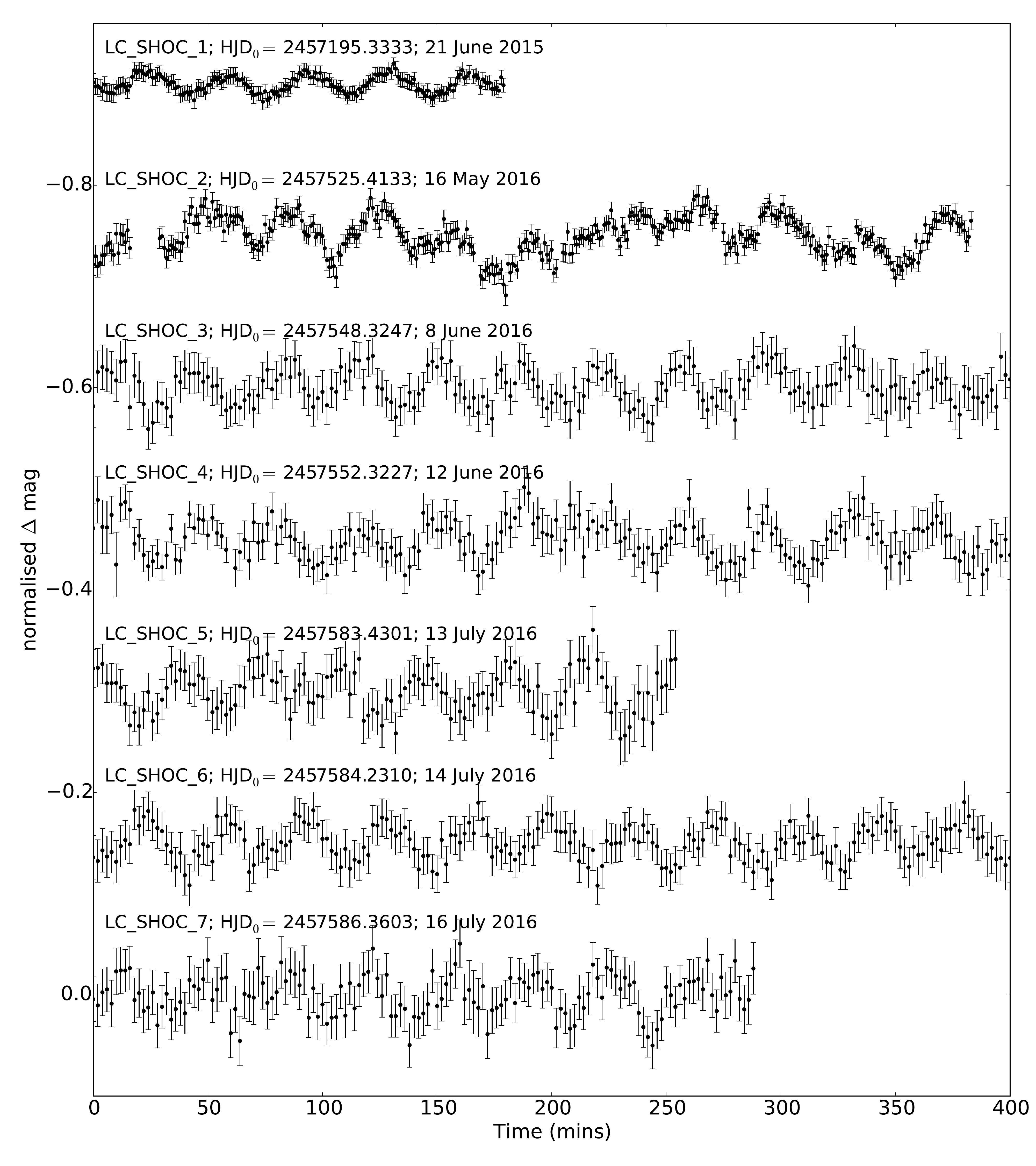}
\caption{All follow-up light curves obtained for OW$\,$J1753--3107, using SHOC cameras on either the SAAO 1.0-m or 1.9-m telescopes. Data points in the light curves are either in 1-min bins (LC$\_$SHOC$\_$1 and LC$\_$SHOC$\_$2) or in 2-min bins (LC$\_$SHOC$\_$3 - LC$\_$SHOC$\_$7), and large trends due to changing airmasses have been removed. light curves have been shifted in the y-axis for visualisation purposes, and are plotted from top to bottom according to their chronological date of observation as labelled (HJD$_{0}$ is the time at the start of each observation). LC$\_$SHOC$\_$1 and LC$\_$SHOC$\_$2 were observed with no filter, and LC$\_$SHOC$\_$3 through to LC$\_$SHOC$\_$7 were observed using a Bessel B filter. See Table~\ref{TAB_PhotObsLog} for the observation log.}
\label{FIG_photlcs}
\end{figure}

Seven high-cadence photometric observing runs of OW$\,$J1753--3107 were conducted over two years: during June 2015, May 2016, June 2016, and July 2016 using the Sutherland High-speed Optical Cameras \citep[SHOC,][]{Coppejans2013} on the 1.0-m and 1.9-m SAAO telescopes, located at the Sutherland Observatory in South Africa. The observing details of these photometric runs are listed in Table~\ref{TAB_PhotObsLog}, and hereafter will be named LC$\_$OW (for the original OW observation), LC$\_$SHOC$\_$1,  LC$\_$SHOC$\_$2, LC$\_$SHOC$\_$3, LC$\_$SHOC$\_$4,  LC$\_$SHOC$\_$5, LC$\_$SHOC$\_$6, and LC$\_$SHOC$\_$7 (in order of observation date, as indicated in Table~\ref{TAB_PhotObsLog}). All SHOC observations were obtained using a 1 MHz read-out speed and a pre-amplifier gain of 2.4 \citep[see][for further details about the SHOC cameras]{Coppejans2013}.

\begin{figure}
\centering
\includegraphics[width=\linewidth]{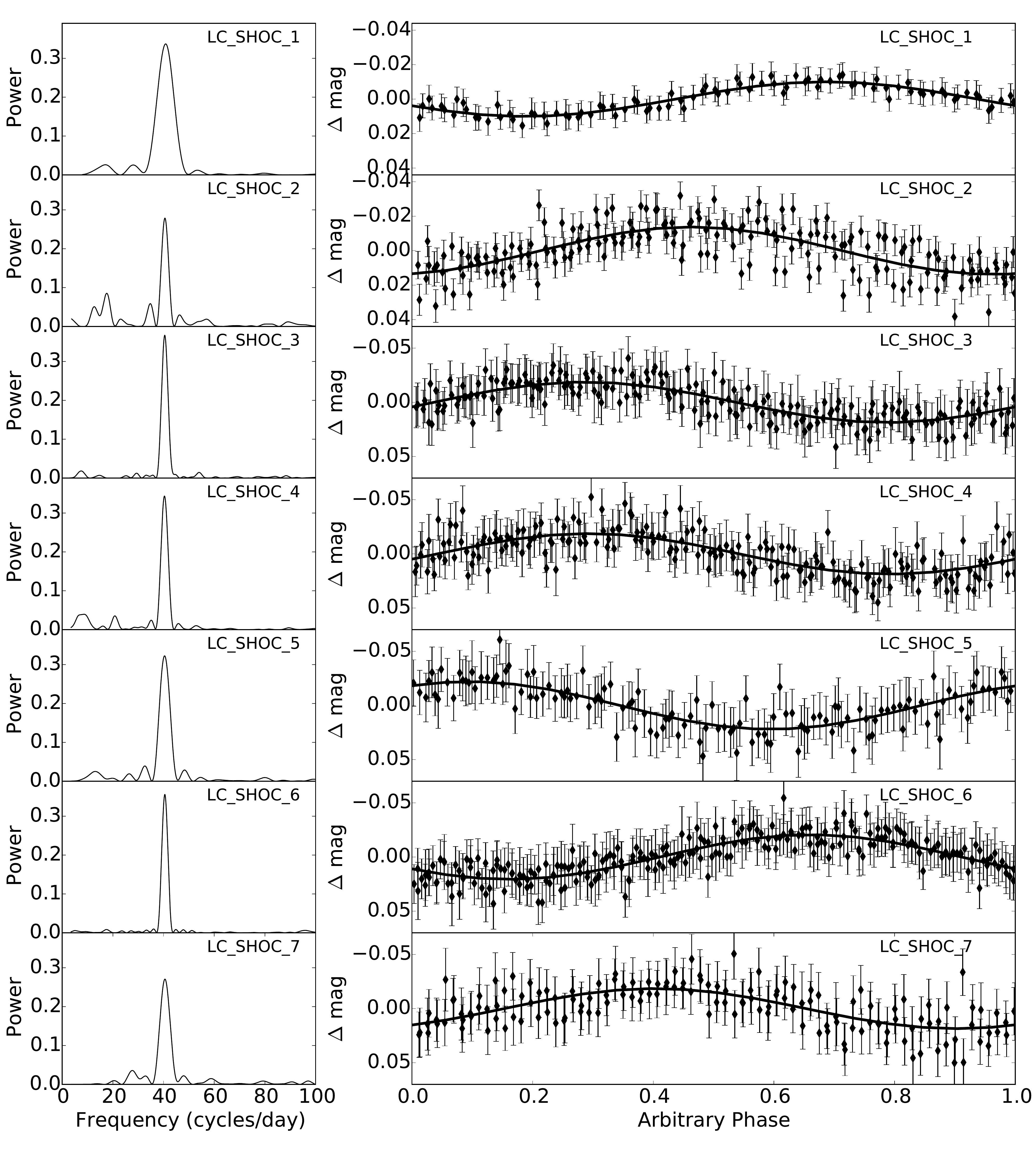}
\caption{Lomb Scargle power spectrum (left panels) and folded light curves (right panels) of the seven follow-up SHOC runs on OW$\,$J1753--3107. In all panels, observing runs are labelled as indicated in Table~\ref{TAB_PhotObsLog}. The power spectrum shows the non-normalised Lomb Scargle periodogram power on the y-axis, and frequency in cycles per day on the x-axis. The light curves (right panels) are folded on the period of the peak power shown in their respective spectrum (and in Table~\ref{TAB_Fitting}), and are in 2-min bins. A single-period sinusoidal model is fit to each of the folded light curves. The data has been observed with either no filter (LC$\_$SHOC$\_$1, LC$\_$SHOC$\_$2) or in B (LC$\_$SHOC$\_$3 - LC$\_$SHOC$\_$7). }
\label{FIG_followup}
\end{figure}

\begin{figure*}
\centering
\includegraphics[width=\linewidth]{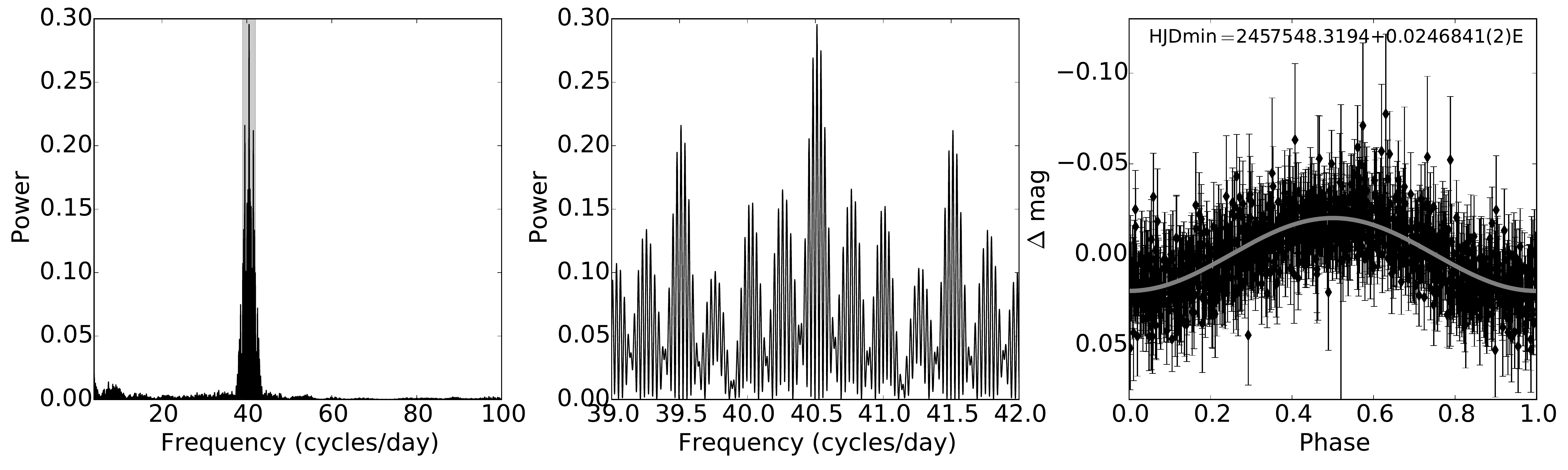}
\caption{Lomb Scargle power spectra (left and middle panels) and phase-folded light curve (right panel) of the combined B-filter SHOC runs on OW$\,$J1753--3107 (LC$\_$SHOC$\_$3 - LC$\_$SHOC$\_$7). The power spectra shows the non-normalised Lomb Scargle periodogram power on the y-axis, and frequency in cycles per day on the x-axis. The spectrum in the middle-panel is a zoomed-in section of the spectrum in the left panel (grey shading). The light curve (right panel, in 2-min bins) has been folded on the ephemeris and period indicated in the figure (HJD$_{min}$), which corresponds to the period of the peak power shown in the spectrum. A single-period sinusoidal model is fit to the folded light curve (grey line).}
\label{FIG_followupcomb}
\end{figure*}

The seven follow-up SHOC light curves are shown in Figure~\ref{FIG_photlcs}, either in 1-min bins (LC$\_$SHOC$\_$1, LC$\_$SHOC$\_$2), or in 2-min bins (LC$\_$SHOC$\_$3 - LC$\_$SHOC$\_$7). The raw data was reduced and light curves extracted using the standard {\tt IRAF} routines \citep{Tody1986,Tody1993}, including bias-subtraction, flat-fielding, and aperture photometry. Using the technique of differential photometry, the average light curve of three bright (stable) neighbouring stars was subtracted from the target light curve in order to compensate for atmospheric effects during the observing runs. Whereas LC$\_$SHOC$\_$1 and LC$\_$SHOC$\_$2 were observed without a filter, the most recent runs, LC$\_$SHOC$\_$3 to LC$\_$SHOC$\_$7 , were observed using a blue filter (Bessell B). Since OW$\,$J1753--3107 is a bluer object than the majority of its neighbouring stars, it can be easily identified in B filter fields. Furthermore, the runs using a B filter were relatively unaffected by the changing airmass throughout the night, unlike the runs observed without a filter which were heavily affected (airmass trends have been removed from the light curves shown in Figure~\ref{FIG_photlcs}). This is due to second order colour terms between the target and the (redder) comparison stars.

From visual inspection of the light curves in Figure~\ref{FIG_photlcs}, the original variations identified in the OW light curve of OW$\,$J1753--3107, fluctuating on a period of $\sim$35 mins, can clearly be seen on all follow-up SHOC light curves. In the case of LC$\_$SHOC$\_$2, the observing conditions on the night of 16 May 2016 were poor, which ultimately injected a false long-period variation of $\sim$174 mins into the light curve (as shown in Figure~\ref{FIG_photlcs}). The non-validity of this period is further confirmed as we do not observe it in the other long duration SHOC light curves. The amplitude of the light curves in LC$\_$SHOC$\_$3 through to LC$\_$SHOC$\_$7 appear to be slightly larger than that of the previous runs, but this is to be expected as the observations were made with two different filters (see Table~\ref{TAB_Fitting}). In order to obtain better estimates on periods present in the light curves, we obtained photometric variability parameters using the {\tt VARTOOLS} package \citep{Hartman2008} on the follow-up observations. These parameters include the period corresponding to the peak Lomb Scargle (LS) Power spectrum (P$_{LS}$). 

A period of $\sim$35 mins corresponding to the peak power is confirmed in the LS power spectrum of each SHOC light curve, as shown in Figure~\ref{FIG_followup} (after the spurious 174-min period is removed from LC$\_$SHOC$\_$2). The LS periods are listed, with errors, in Table~\ref{TAB_Fitting}. We do not find any significant peaks above 100 cycles/day and, although there are a few other smaller peaks in the power spectrum, they are not featured in every spectrum as is the case of the overwhelming peak period. Thus, we conclude that OW$\,$J1753--3107 \textcolor{black}{has a single photometric period, equal to $\sim$ 35 mins}. Furthermore, this period has remained approximately constant from the first OW observation in July 2014, to two years later with the LC$\_$SHOC$\_$7 run in July 2016. 

In order to test the stability and accuracy of the single-period assumption, we folded each light curve on their respective P$_{LS}$, as shown in Figure~\ref{FIG_followup} in 2-min bins for visualisation purposes. We fit each folded light curve with a single-mode sinusoidal curve using parameters determined through {\tt VARTOOLS}, and a simple sinusoidal model of the form:
\[
f(\rm t) = A cos(\omega t + \psi)                                
\]

where $f(\rm t)$ is the model light curve flux as a function of time t, $\omega$ is the frequency associated with the peak LS periodogram power, $A$ is the amplitude of the model light curve, and $\psi$ gives the phase of the rotation at the moment of observation. 

The fitting parameters for the light curve of each observing run (including LC$\_$OW) are listed in Table~\ref{TAB_Fitting}. In all SHOC light curves, the folded data agree well with a single-mode sinusoidal model fit, having median residual values between 0.2 $\%$ and 1.6 $\%$ (see Table~\ref{TAB_Fitting}). The unexpected scatter seen in LC$\_$SHOC$\_$2 is due to poor atmospheric conditions during the observing run. The ephemeris is found to be HJD$_{min}$ $=$ 2457548.31942 $+$ 0.0246841(2) E. Furthermore, by combining the five B filter runs (LC$\_$SHOC$\_$3 - LC$\_$SHOC$\_$7), we find OW$\,$J1753--3107 exhibits a single-mode frequency of 4.68887 (3) $\times$ 10$^{-4}$ Hz, corresponding to 35.5452 (2) min. This is further shown in the folded light curve and power spectrum of the combined B filter runs in Figure~\ref{FIG_followupcomb}.

\begin{table}
\centering
\caption{Peak Lomb Scargle periodogram period (P$_{LS}$) with error, and sinusoidal fitting parameters$^{*}$ for all photometric runs on OW$\,$J1753--3107, including the peak frequency in the power spectrum ($f$), the amplitude of the peak frequency in the power spectrum ($A$), and the phase of the fitted light curve ($\psi$). Also shown is the median residual of the model fit to the folded light curve data (resid$_{median}$), shown in Figure~\ref{FIG_followup}.}
\label{TAB_Fitting}
\resizebox{0.48\textwidth}{!}{
\begin{tabular}{lccccc}
\hline
RunID      & P$_{LS}$        & $f$      & $A$       & $\psi$  & resid$_{median}$ \\
           & (mins)       &  (cycles/day)   & ($\%$)   & (rad)     &     ($\%$)             \\
\hline           
LC$\_$OW     & 35.3  $\pm$  3.9  & 0.0245 & 1.3 & 0.09  & 0.5     \\
LC$\_$SHOC$\_$1 & 35.3  $\pm$ 3.3  & 0.0245 & 1.0  & --1.18  & 0.2  \\
LC$\_$SHOC$\_$2 & 35.5  $\pm$ 1.4  & 0.0247 & 1.4 & 0.23  & 0.9     \\
LC$\_$SHOC$\_$3 & 35.6  $\pm$ 1.3  & 0.0247 & 1.9 & 1.34  & 1.6     \\
LC$\_$SHOC$\_$4 & 35.7  $\pm$ 1.4  & 0.0248 & 1.9 & 1.30  & 1.0      \\
LC$\_$SHOC$\_$5 & 35.6  $\pm$ 2.1  & 0.0247 & 2.2 & 2.55  & 0.7      \\
LC$\_$SHOC$\_$6 & 35.6  $\pm$ 1.2  & 0.0247 & 2.1 & 2.13 & 0.7    \\
LC$\_$SHOC$\_$7 & 35.5  $\pm$ 2.0  & 0.0247 & 1.9 & 0.61  & 1.0     \\
\hline
\multicolumn{6}{p{8.0cm}}{$^{*}$sinusoidal model fit, see Section~\ref{SEC:Photometry} for model description.}\\
\end{tabular}}
\end{table}

\section{Spectroscopic Analysis}
\label{SEC:Spectroscopy}

The spectrum of OW$\,$J1753--3107, as shown in Figure~\ref{FIG_discovery} and discussed in Section~\ref{SEC:Discovery}, reveals clear evidence for carbon and helium, which is very similar in appearance to the spectrum exhibited by the first warm DQ SDSS$\,$J1036$+$6522 \citep{Williams2013}. Hence, we are able to fit the spectral features exhibited by OW$\,$J1753--3107 using an adaptation of the model that was applied to SDSS$\,$J1036$+$6522 by \citet{Williams2013}. Fitting the spectral features allow us to obtain better estimates of the physical properties of OW$\,$J1753--3107, including the temperature ($T_{\rm eff}$), surface gravity ($g$), magnetic field strength (B$_{z}$), and carbon abundance (N(C)/N(He)). Furthermore, we obtained high resolution spectra for precise spectral line identification purposes. Details of the three spectroscopic follow-up observing runs are shown in Table~\ref{TAB_SpecObsLog}, and are discussed in detail in Section 4.1 and Section 4.2. 

\begin{table*}
\centering
\caption{Spectroscopic observing log, including the following parameters: date of observation (Date Obs.), time at the start of the observation (HJD$_{0}$), the telescope (Tel.), instrument (Instr.) and mode used, the grating (Gr.) and grating angle (Gr.  Angle) used, the slit width, the length of each exposure (Exp. Time), and the dispersion across the wavelength range (Disp.)}
\label{TAB_SpecObsLog}
\resizebox{\textwidth}{!}{
\begin{tabular}{cccccccccc}
\hline
Date  & HJD$_{0}$           & Tel.  & Instr. & Mode      & Gr. & Gr.  & Slit  & Exp. & Disp. \\
Obs. &             &   &  &      &  & Angle & Width & Time  &             \\
(dd/mm/yyyy)  &                  &            &            &           &         & ($\degr$)   & ($\arcsec$)  & (s)     & (\AA/pixel)                        \\
\hline
30/04/2016    & 2457509.4982    & SAAO 1.9-m & SpUpNIC    & longslit  & gr7     & 17.4      & 1.8        & 2 $\times$ 1800 & 2.80\\
15/05/2016    & 2457540.4526 & SALT       & RSS        & longslit  & pg1300  & 18.88    & 1.5        & 3 $\times$ 720  & 0.66  \\
01/06/2016    & 2457540.9525    & Keck II    & ESI        & echellete &         &           & 1          & 7 $\times$ 240, 12 $\times$ 180 &  0.16 \\
\hline
\end{tabular}}
\end{table*}

\subsection{Spectral Line Identification}

\begin{figure*}
\centering
\subfloat{\includegraphics[width=\textwidth]{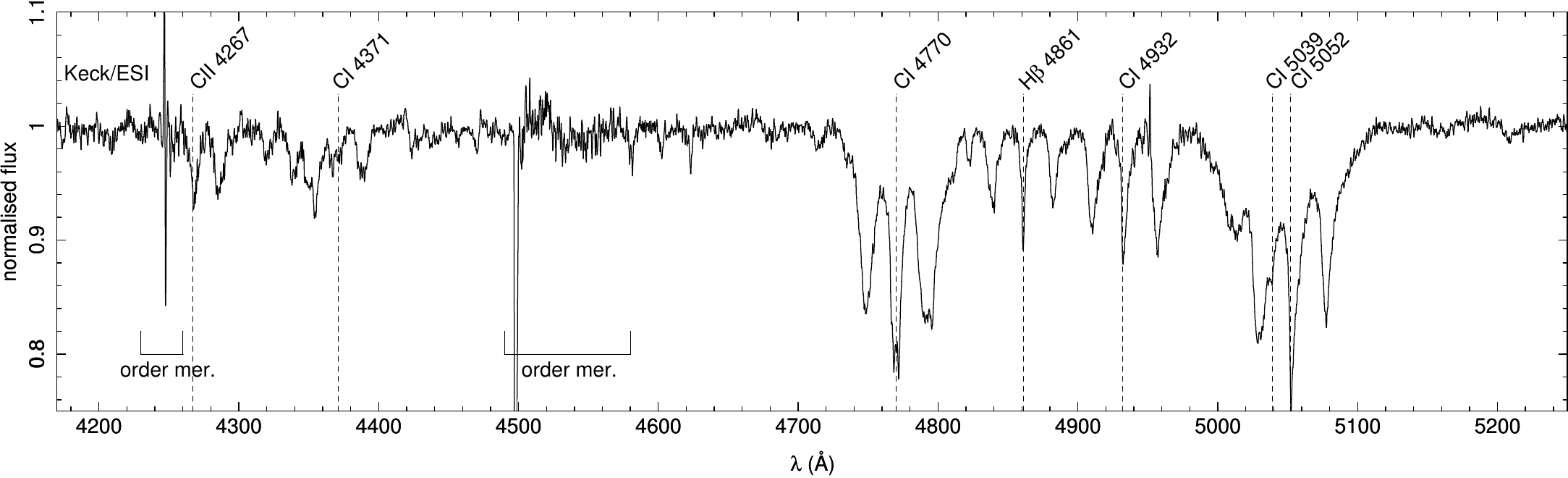}}
\hspace{0.2cm}
\subfloat{\includegraphics[width=\textwidth]{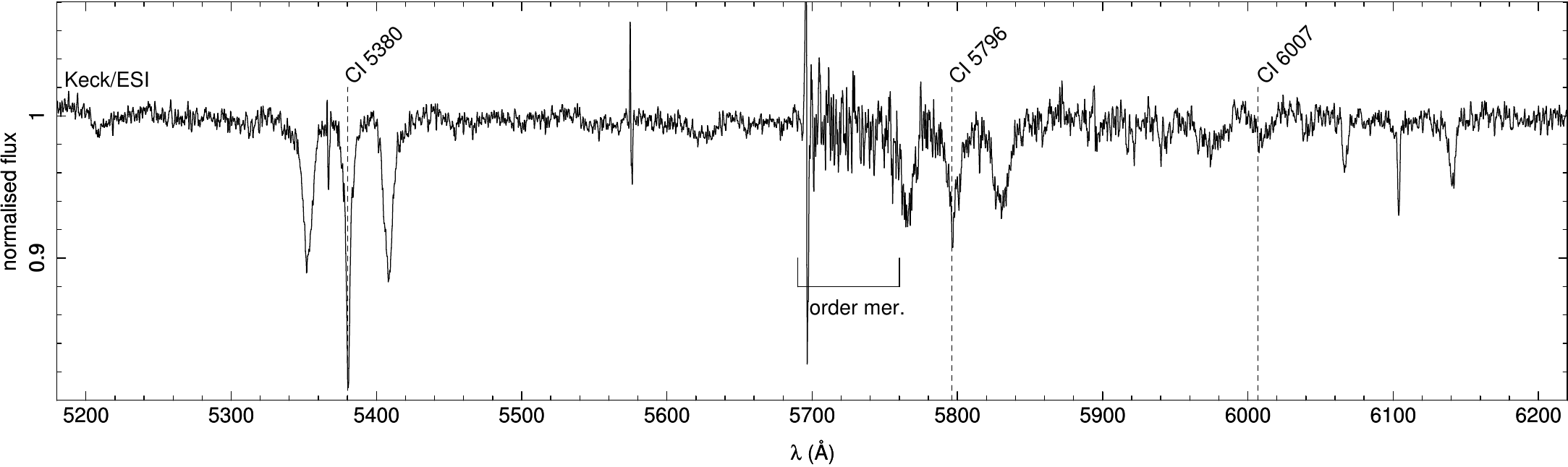}} 
\hspace{0.2cm}
\subfloat{\includegraphics[width=\textwidth]{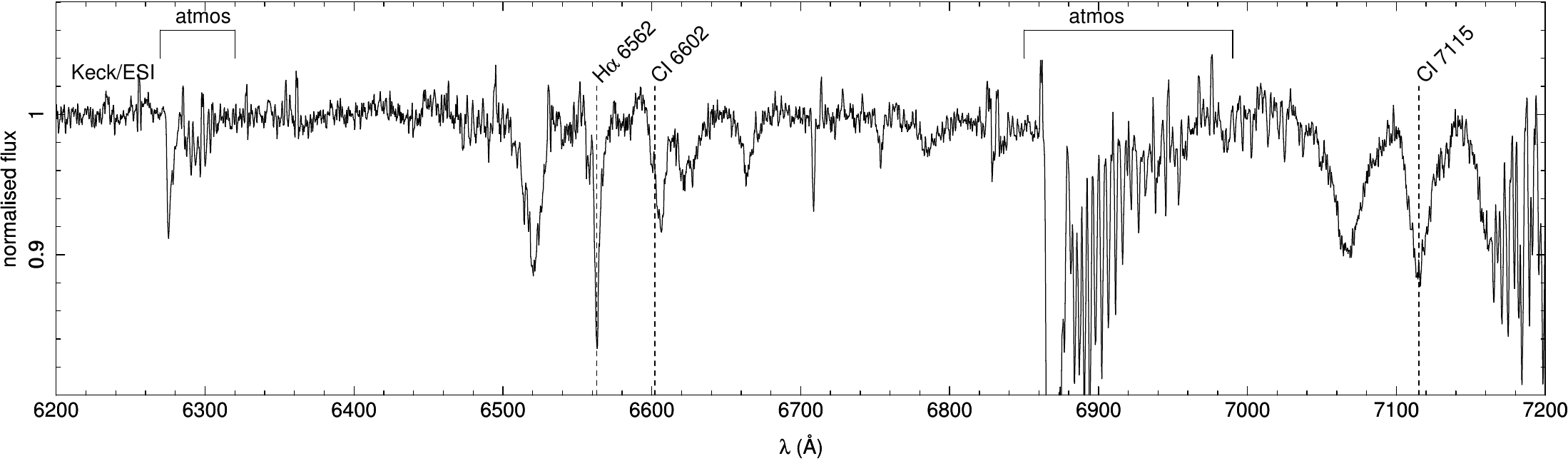}} 
\caption{Average spectrum of OW$\,$J1753--3107, observed using the echelle mode on the ESI instrument on the Keck telescope. The spectrum has been Gaussian-smoothed and is continuum-normalised. Spectral lines have been labelled. Regions of echelle order mergers (order mer.) are indicated. }
\label{FIG_Keckspectrum}
\end{figure*}

In order to precisely identify any absorption lines shown in the spectrum of OW$\,$J1753--3107 and to search for radial velocity shifts, we obtained 18 3-min or 4-min spectra using the Echelette Spectrograph and Imager instrument \citep[ESI,][]{Sheinis2002} located on the Keck II telescope. Exposures were taken in echellette mode, with a slitwidth of 1\arcsec, and the spectra were reduced and extracted using the {\tt MAKEE}\footnote{http://www.astro.caltech.edu/$\sim$tb/makee/} pipeline. A wavelength solution was found for the spectra by calibrating with HgNeXeCuAr arcs, and instrumental flexure was corrected for using CuAr arcs observed at the beginning of the night.      

The average-combined and continuum-normalised spectrum is shown in Figure~\ref{FIG_Keckspectrum}, with prominent absorption lines identified and labelled; \textcolor{black}{the absorption line at $\sim$6105 {\AA} remains unidentified.} The spectrum is the combination of the 10 orders on the CCD, and span the wavelength range from 4170$\AA$ -- 7200$\AA$. The combination of the orders can cause relatively higher noise levels where they have been stitched together, as shown in the region around $\sim$4500 \AA. In the spectrum, we are able to make out clear Zeeman-split lines of C I and C II. There is also possible evidence for the H$\alpha$ and H$\beta$ lines identified in the SpUpNIC spectrum (Figure~\ref{FIG_discovery}). However, there is no apparent evidence for the He I $\lambda$4471 broad absorption dip that was present in the SpUpNIC spectrum. This is likely a consequence of the continuum fit to the individual orders. 

\begin{table}
\centering
\caption{Radial velocity measurements for spectral lines of H$\alpha$, H$\beta$, and C\,I $\lambda$5380 using {\tt FITSB2} \citep{Napiwotzki2004} in the 18 KECK ESI spectra.}
\label{TAB_radvel}
\begin{tabular}{cr}
\hline
Heliocentric Julian Date & Velocity (km/s)\\
\hline
245\,7540.9593649	&	70.5$\pm$6.7	\\
245\,7540.9627950	&	68.0$\pm$5.5	\\
245\,7540.9661938	&	77.0$\pm$6.3	\\
245\,7540.9696060	&	66.4$\pm$6.4	\\
245\,7540.9730071	&	70.8$\pm$8.3	\\
245\,7540.9764268	&	79.8$\pm$8.6	\\
245\,7540.9798291	&	82.6$\pm$7.0	\\
245\,7540.9829443	&	82.7$\pm$9.3	\\
245\,7540.9857036	&	84.3$\pm$15.8	\\
245\,7540.9883721	&	73.5$\pm$8.2	\\
245\,7541.0014171	&	68.3$\pm$23.6	\\
245\,7541.0042447	&	91.1$\pm$8.9	\\
245\,7541.0068692	&	83.0$\pm$8.7	\\
245\,7541.0095776	&	70.7$\pm$9.3	\\
245\,7541.0124886	&	72.2$\pm$14.1	\\
245\,7541.0226497	&	87.3$\pm$16.0	\\
245\,7541.0253813	&	81.4$\pm$12.3	\\
245\,7541.0280880	&	75.3$\pm$11.4	\\
\hline
\end{tabular}
\end{table}  

\begin{figure}
\centering
\includegraphics[angle=270,origin=c,width=\linewidth]{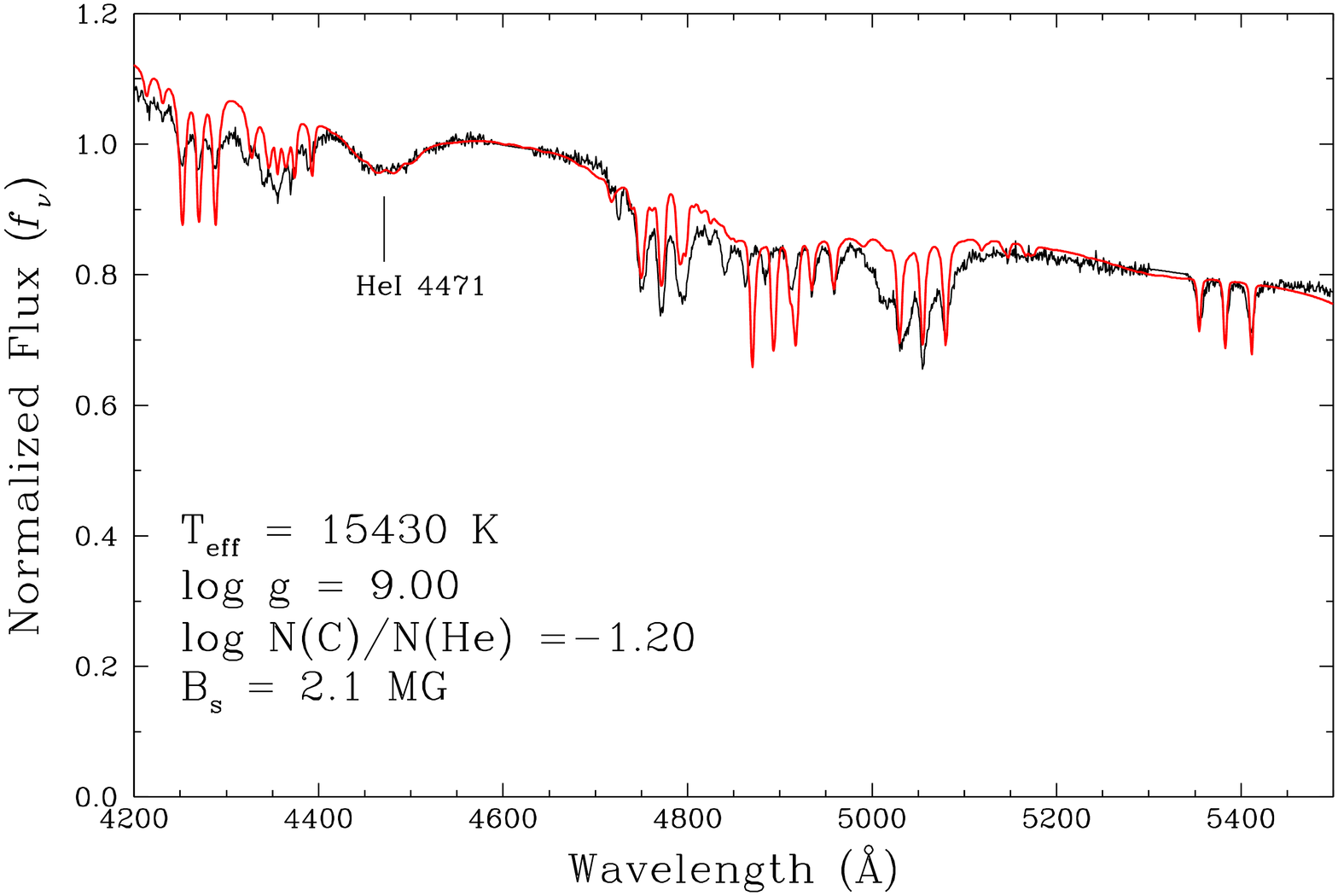} \vspace{-2cm}
\caption{Average-combined and normalised SALT spectrum of OW$\,$J1753--3107 (black line), with the best-fit model spectra overlain (red line). The physical properties, including the temperature ($T_{\rm eff}$), surface gravity ($g$), magnetic field strength (B$_{z}$), and carbon abundance (N(C)/N(He)) determined from the model fit are indicated. The model line strengths of the central C\,I lines around 4900 \AA, 4340 \AA\ and He\,I $\lambda$4471 are in agreement with the spectral data. However the model does not fit several large-scale dips across the spectrum which are possibly due to spectral line identification issues.}
\label{FIG_specfit}
\end{figure}

In the 18 KECK ESI spectra, radial velocities were measured by fitting a Lorentzian/Gaussian function to the spectral lines of H$\alpha$, H$\beta$, and C\,I $\lambda$5380 using {\tt FITSB2} \citep{Napiwotzki2004}. These radial velocities are shown in Table~\ref{TAB_radvel}, where the large error margins are the result of the low SNR in the single spectra, and the lack of strong unblended lines to measure. We were not able to detect any significant radial velocity variation in the spectral lines \textcolor{black}{at the 35-min period to an amplitude limit of $<$ 5 km/s, which} suggests that OW$\,$J1753--3107 is not a UCB system.    \\  

\subsection{Spectral line fitting}

For spectral modelling, we obtained spectra of OW$\,$J1753--3107 using the Robert Stobie Spectrograph \citep[RSS,][]{Kobulnicky2003} on the 10-m Southern African Large Telescope \citep[SALT][]{Buckley2006}. As shown in the spectroscopic observing log in Table~\ref{SEC:Spectroscopy}, three 12-min exposures were made using the RSS in long-slit mode, with a 1.5\arcsec\ slit width. Furthermore, the grating pg1300 was used with a grating angle of 18.88\degr, providing a wavelength range of 3898 \AA\ - 5981 \AA\ with a wavelength sampling of $\sim$1.5 pixels/\AA. The spectra was processed and extracted using standard {\tt IRAF} \citep{Tody1986,Tody1993} and \textsc{PYSALT} \citep{Crawford2014} routines, and {\tt L.A.Cosmic} software \citep[for cosmic ray reductions,][]{Dokkum2001}. A CuAr arc was used for wavelength calibration, and a spectrum of the standard star LTT 4364 was compared with for flux calibration. 

The average combined SALT spectrum is shown in Figure~\ref{FIG_specfit}. To this spectrum, we fit an adaptation of the LTE model atmospheres described in detail in \citet{Williams2013}, which uses line splitting calculated in the Paschen-Back regime to form a grid of model atmospheres that consist of both helium and carbon. The model makes the assumptions that a) the magnetic field does not affect the thermodynamic structure, b) the target is undergoing normal radiative transfer, and c) B$_{z}$ is uniform over the surface of the star. See  \citet{Williams2013} for a detailed description of this model.

Using a non-linear least-squares routine, a best-fit solution is found for the spectra, as shown in  Figure~\ref{FIG_specfit}. The solution is degenerate, and thus a change in one parameter ($T_{eff}$, $g$, or N(C)/N(He)) can be compensated for by a change in the other parameters. From visual inspection, the model manages to correctly replicate the line strengths of almost all the spectral features within this wavelength range. However, there appears to be large scale discrepancies between the spectral data and the model fit (around the 4340 \AA\ and 4900 \AA\ regions), which is possibly due to spectral line identification issues. Furthermore, the model does not take into account any hydrogen in the atmosphere, and thus the possible H$\beta$ absorption line (at 4861 \AA) has not been fit. The values of $T_{\rm eff}$, $g$, B$_{z}$, and N(C)/N(He), used to formulated the best fit model, are found as follows:

\begin{itemize}
\item A magnetic field strength of B$_{z} =$ 2.1 MG is determined by measuring the separation between the line components of the magnetically split C\,I triplet component at 5380 \AA. 
\item The broad He\,I $\lambda$4471 dip is used to calculate the carbon abundance, where $\log{(N(C)/N(He))} = $ \textcolor{black}{--1.2}
\item The best fit model solution gives a temperature of $T_{\rm eff} =$ \textcolor{black}{15430} K, and a surface gravity of  $\log{(g)} =$ \textcolor{black}{9.0}. The $T_{\rm eff}$ is determined from fitting the photometry (with $\log{(g)}$ fixed at \textcolor{black}{9.0} and $\log{(N(C)/N(He))}$ determined from the spectra).   
\end{itemize}

\begin{figure}
\centering
\includegraphics[angle=270,origin=c,width=\linewidth]{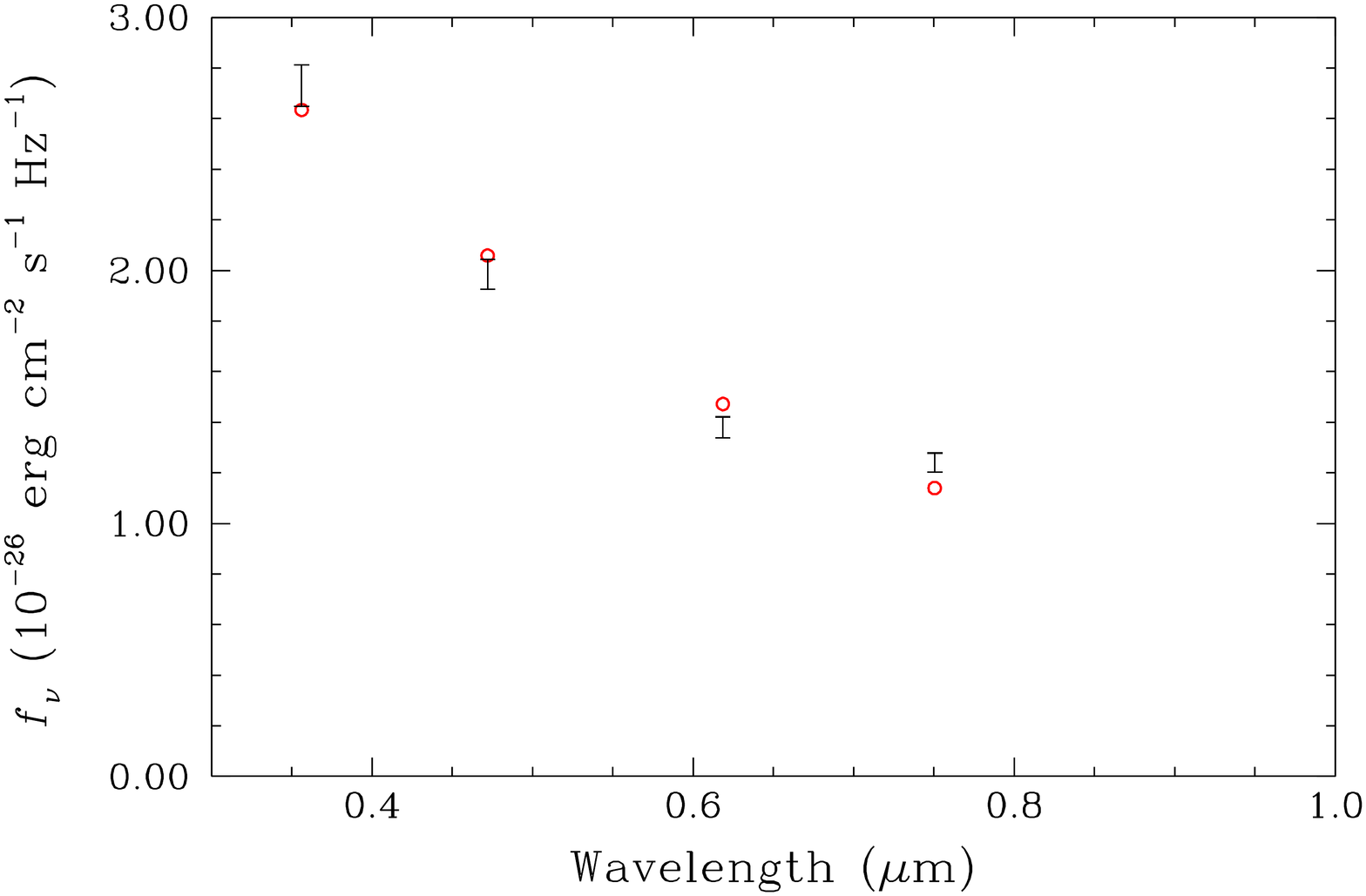} \vspace{-2cm}
\caption{Comparison of the average fluxes (red circles) obtained from the best fit spectral model, and the fluxes from $u-$,$g-$, $r-$ and $i-$band VPHAS+  colours (error bars) for the DQ OW$\,$J1753--3107. The best-fit model is shown in Figure~\ref{FIG_specfit}.}
\label{FIG_specdist}
\end{figure}

Although the line strengths are generally well reproduced by the model, further modelling of improved spectra should provide more accurate estimates of the physical properties, which in turn can be used to find the mass of the target. In Figure~\ref{FIG_specdist}, we compare the VPHAS+ photometry obtained for OW$\,$J1753--3107 with photometry calculated from filter response curves folded to the best-fit model spectra. This shows us that the model data agrees with the VPHAS+ photometry to within 3$\sigma$ error bars.

\section{Discussion}

The photometric and spectroscopic features of OW$\,$J1753--3107 reveal it is a warm carbon-enriched DQ. Indeed, it can be considered a twin of the first warm DQ SDSS$\,$J1036$+$6522, but varying on a longer period and with a higher carbon abundance (see Table~\ref{TAB:warmDQVcomp} for a comparison of the properties between these two warm DQ). In the following sections, we look at the similarities and differences between these two warm DQ, and discuss some key questions in hot and warm DQ research, namely a) what is the nature of OW$\,$J1753--3107, and what does this imply for current theorised evolutionary channels?, b) what is the origin of the observed variations, and c) what does the magnetic nature of the system imply?

\begin{table}
\centering
\caption{Comparison of the physical properties of the two warm DQ: OW$\,$J1753--3107 and SDSS$\,$J1036$+$6522. Properties include: the effective temperature ($T_{\rm eff}$), surface gravity ($\log{(g)}$), magnetic field strength (B$_{z}$), carbon abundance ($\log{(N(C)/N(He))}$), period (P) and amplitude ($A$) of mono-periodic variations, and $g$-band magnitude (in Vega colour system).}
\begin{tabular}[pos]{lll}
\hline
Property  & OW$\,$J1753--3107 & SDSS$\,$J1036$+$6522\\
\hline
$T_{\rm eff}$ (K) & \textcolor{black}{15430} & 15500\\
$\log{(g)}$ & \textcolor{black}{9.0} & 9.0\\
$\log{(N(C)/N(He))}$ & \textcolor{black}{--1.2} & --1.0\\
B$_{z}$ (MG) & 2.1 & 3.0\\
P (mins) & 35.5 & 18.6 \\
$A$ ($\%$) & 1.3 & 0.4\\ 
$g$ (mag) & 15.74 & 18.58\\
\hline
\end{tabular}
\label{TAB:warmDQVcomp}
\end{table}

\subsection{Nature of the system}

The high abundance of carbon derived from the modelled atmospheres of hot DQs, along with such a constrained temperature range can be explained by the progenitor and evolutionary channel model proposed by \citet{Dufour2008a}. The carbon abundance observed in these sources are akin to those seen in massive PG1159 stars, such as H1054$+$65 \citep[e.g.,][]{Werner2004} and RX J0439.8-6809 \citep{Werner2015}. Both of these massive WD are extremely hot ($T_{\rm eff}$ $>$ 200000 K) and consist of atmospheres rich in carbon and oxygen. Their hydrogen and helium-deficient atmospheres are believed to be the result of a very late and violent thermal pulse during the post-asymptotic-giant-branch phase which destroys most of the H/He envelope. As the residual wind from this violent phase fades, gravitational settling causes the remaining helium, carbon and oxygen to separate, and the helium to form a thin shell surrounding the carbon-enriched mantle. The He atmosphere WD (with spectral type DB, $T_{\rm eff}$ $\approx$ 25000 K) then cools down further and a convection zone in the carbon layer forms, causing the carbon to start to diffuse into the outer helium layer. The object is now a carbon-enriched hot DQ WD, with a temperature that is in the 18000 K to 24000 K range and a carbon abundance of $\log{(N(C)/N(He))} >$ 1. The hot DQ will then continue to cool as helium begins to resurface and, thus, it will transition via a ``warm" DQ phase into a second sequence of helium-enriched cool DQ WD. 

\begin{figure}
\centering
\includegraphics[width=\linewidth]{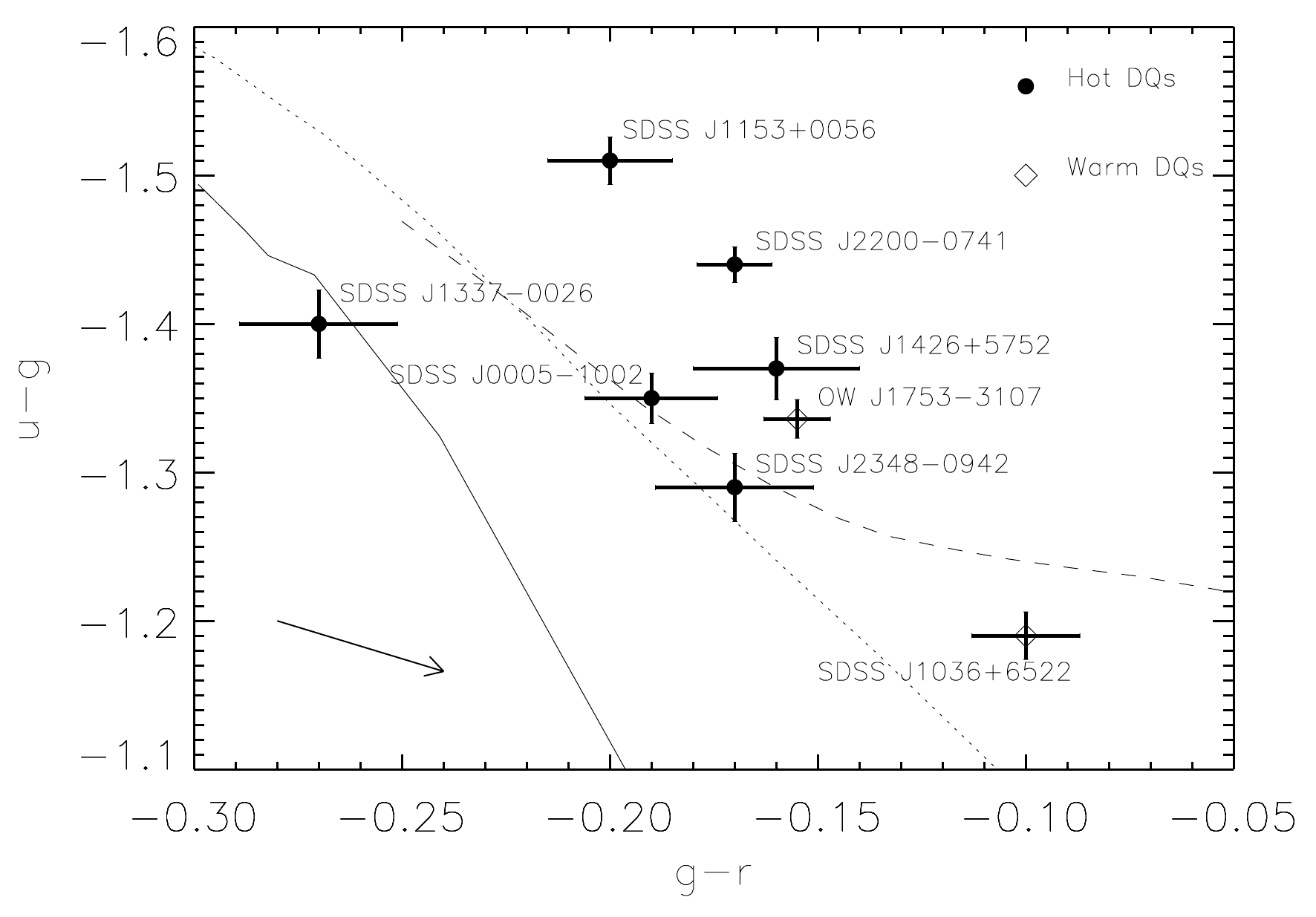}
\caption{Distribution of all eight warm and hot DQs in $u-g$ vs $g-r$ space. We overlay the Main Sequence (MS) track (solid line), created using synthetic colours provided by \citet{Drew2014} for R$_v$=3.1 and extinction coefficient A$_0$ = 0, indicated by `Drew MS'. DA and DB tracks are shown, as dotted and dashed lines, respectively, using synthetic colour colour tracks with surface gravities of $\log{(g)} =$ 8.0 provided by \citet{Raddi2016}. All colours are in the standard (Vega) system, and are listed in Table~\ref{TAB:DQVcolourcomp}. All eight DQ, including our new discovery OW$\,$J1753--3107, are labelled. The reddening vector displayed in the bottom left of this figure represents A$_V$ = 0.1.}
\label{FIG_colcol}
\end{figure}

The relatively low temperature and carbon abundance of OW$\,$J1753--3107 suggest it indeed resides in this transition state. Similarly to SDSS$\,$J1036$+$6522, the temperature of OW$\,$J1753--3107 ($T_{\rm eff} =$ \textcolor{black}{15430} K) is lower than the expected temperature range of hot DQs. In Table~\ref{TAB:DQVcolourcomp}, we compare the colours (in the Vega magnitude system\footnote{Colours of the SDSS hot DQs have been transformed from the AB magnitude system into the Vega System using the equations of \citet{Blanton2007}. We note that there appears to be some uncertainty between the conversion from SDSS AB colours and VPHAS+
  colours at the level of a few \textcolor{black}{hundredths} of mag in the $r$-band, rising to more than one tenth of a mag in the $u$-band, \textcolor{black}{see \citet{Drew2014} for details}.}) of OW$\,$J1753--3107 to other known warm or hot DQs, and show their location in colour-colour space in Figure~\ref{FIG_colcol}. Firstly, we can see from this table, that OW$\,$J1753--3107 is brighter by more than 2 magnitude in $u$, $g$, $r$ and $i$ colour bands. Secondly, the $u-g$ and $g-r$ colours of OW$\,$J1753--3107 appears consistent with other warm or hot DQs. This can also be seen in Figure~\ref{FIG_colcol}. Although the $r-i$ of our target appears relatively higher, this is possibly due to Galactic reddening as OW$\,$J1753--3107 is located towards the Galactic Bulge. \textcolor{black}{On the other hand, we deduce a photometric distance of 40.5$_{-2.7}^{+3.4}$ pc (assuming $\log{(g)}$ = 9.0), so the Galactic reddening may be small or negligible.}
\textcolor{black}{We searched for evidence for proper motion using the SuperCOSMOS images \citet{Hambly2001a}. There is a clear offset in the position
of OW 1753-3107 between the OW epoch (2014 Jul 29) and the SuperCOSMOS blue image (1974 Jun 17). The proper motion derived from SuperCOSMOS data for OW$\,$J1753--3107 is $-0.7$ mas/yr in RA and $-$136 mas/yr in DEC \citet{Hambly2001b}. The {\sl Gaia} Data Release 2 schedulded for April 2018 will include the parallax and proper motion of OW$\,$J1753--3107 and will allow us to determine whether it is a member of the Galactic disk or halo.}

From Figure~\ref{FIG_colcol}, it appears that the known sample of warm and hot DQs are located in a specific region of colour-colour space, implying that colour cuts can be applied to surveys such as OW in order to improve the future detection of new warm or hot DQs. However, with the current sample of these DQ, we are not yet able to confirm if warm and hot DQs cannot be differentiated based on their location in colour-colour space.

\begin{table*}
\centering
\caption{Colour properties of the eight warm and hot DQs. The colours of the previously known warm and hot DQ variables have been converted to lie on the Vega colour scale.}
\begin{tabular}[pos]{lccccccc}
\hline
Star ID & $u$     & $g$     & $r$     & $i$          & $u-g$ & $g-r$   & $r-i$ \\
		& (mag)   & (mag)   & (mag)   & (mag)     & (mag) & (mag) & (mag)\\
\hline
\multicolumn{8}{l}{Hot DQs}\\
\hline
SDSS$\,$J142625.70$+$575218.4 & 17.85 & 19.22 & 19.38 & 19.59 & --1.37 & --0.16 & --0.21 \\
SDSS$\,$J220029.08--074121.5 & 16.39 & 17.82 & 18.00 & 18.11 & --1.44 & --0.17 & --0.11 \\
SDSS$\,$J234843.30--094245.3 & 17.79 & 19.09 & 19.25 & 19.35 & --1.29 & --0.17 & --0.10 \\
SDSS$\,$J133710.19--002643.6 & 17.31 & 18.71 & 18.98 & 19.07 & --1.40 & --0.27 & --0.09 \\
SDSS$\,$J115305.54$+$005646.2 & 17.47 & 18.97 & 19.17 & 19.28 & --1.51 & --0.20 & --0.11 \\
SDSS$\,$J000555.90--100213.5 & 16.42 & 17.77 & 17.96 & 18.10 & --1.35 & --0.19 & --0.15 \\
\hline
\multicolumn{8}{l}{Warm DQ}\\
\hline      
SDSS$\,$J103655.38$+$652252.0 & 17.39 & 18.58 & 18.68 & 18.78 & --1.19 & --0.10 & --0.10 \\
OW$\,$J175358.85--310728.9   & 14.40 & 15.74 & 15.89 & 15.80 & --1.34 & --0.16 & 0.09 \\
\hline
\end{tabular}
\label{TAB:DQVcolourcomp}
\end{table*}

\subsection{The Origin of Variations}

One of the leading questions in hot DQ research concerns the origin of the observed variations. Early predictions by \citet{Montgomery2008} and  \citet{Fontaine2008} suggested that some hot DQs would be unstable to non-radial pulsations, which lead to the first discovery of variability in a hot DQ WD in SDSS$\,$J1425$+$5652 \citep[][]{Montgomery2008}.  The variations observed in the first five hot DQs, all with periods under 20 minutes, were originally attributed to non-radial pulsations \citep{Dufour2009,Dufour2011}, perhaps in analogy to large-amplitude rapidly oscillating A stars . However, the detection of a longer-period variation of 2.1 days in the sixth discovered hot DQ SDSS$\,$J0005--1002 (with no evidence of a shorter period) was attributed to the rotation of a magnetic white dwarf \citep{Lawrie2013}. Indeed, rapid rotation of the WD has now been proposed as the leading theory to explain the observed variations in hot DQs \citep{Williams2016}, supported by the apparent evidence that most hot DQs only exhibit single-mode variations (with related harmonics), as opposed to the multi-mode variations observed in pulsating WD. Although rotational periods are typically on the scale of hours to days, rotational periods as short as 10s of minutes have been observed in some white dwarfs \citep[for e.g., the 12-min rotations observed in the white dwarf RE J0317-853,][]{Barstow1995}.  

Since this rotation theory has be been used to explain the 18.6-min variations seen in the warm DQ SDSS$\,$J1036$+$6522 \citep{Williams2013}, we can similarly use it as an explanation for the 35-min variations observed in OW$\,$J1753--3107. However, as both warm DQ are at a lower temperature than hot DQs and are thus theorised to be in a transition state, the source of the variations may be different to that of the other hot DQs (as suggested by \citet{Williams2013} for the variations observed in SDSS$\,$J1036$+$6522). Furthermore, unlike the short-period (under 20 minute) or 2.1 day variations of other hot DQs, OW$\,$J1753--3107 exhibits variations on a single period of $\sim$35 mins, with no evidence for other shorter or longer periods. Thus, it is likely the variations are not due to pulsations, but rather to the rotation of a magnetic white dwarf.

\subsection{Magnetic behaviour}
Close to 70$\%$ of hot DQs have been found to be strongly magnetic with magnetic field strengths B$_{z} >$ 1 MG \citep{Dufour2011,Williams2016}. The magnetic nature of some of the warm and hot DQs is also evident in the Zeeman splitting of their spectral lines (as in OW$\,$J1753--3107, see Figure~\ref{FIG_discovery}). By modelling these spectral features and measuring the degree of separation in the split carbon lines, \citet{Williams2013} were able to detect a magnetic field strength in SDSS$\,$J1036$+$6522 of 3.0 $\pm$ 0.2 MG. Similarly, the Zeeman split carbon lines exhibited by OW$\,$J1753--3107 were used to calculate a magnetic field strength of B$_{z} =$ 2.1 MG.    

It has been previously suggested that the observed non-linearities in the light curve pulse shape of some hot DQs may be related to the presence of a strong magnetic field \citep[e.g. ][]{Green2009}. For example, the non-magnetic hot DQ SDSS$\,$J2348--0942 exhibits a sinusoidal pulse shape, whereas the hot DQ SDSS$\,$J2200--0741 exhibits non-sinusoidal variations (due to the presence of harmonics) and is strongly magnetic. However, SDSS$\,$J1036$+$6522 contradicts this theory, as it is strongly magnetic, yet exhibits sinusoidal pulse shapes \citep{Williams2016}. As the photometric analysis in this paper shows, OW$\,$J1753--3107 is also highly magnetic, and similarly shows a clear sinusoidal pulse shape, unchanging over 2 years. Thus, this is further evidence against a non-linear pulse shape and magnetic field strength relationship.

This model of magneto-rotational variations can be addressed using high-resolution phase-resolved spectropolarimetry. We expect the degree of linear polarisation to strongly change when the magnetic field poles rotate into/out of the line of sight.

\section{Conclusions}
In this paper, we present details of a new warm DQ OW$\,$J1753--3107, discovered in the OmegaWhite Survey. Here we summarise the main results of our photometric and spectroscopic analysis:
\begin{itemize}
 \item OW$\,$J1753--3107 exhibits variations on a dominant period of 35.5452 (2) mins, with no clear evidence for harmonics or other non-related periods. A single-sinusoidal model provided a good fit for all follow-up light curves. Furthermore, the 35-min period has remained constant over a 2-year span from 2014 to 2016.
 \item The follow-up spectra of OW$\,$J1753--3107 reveal strong evidence for C\,I and C\,II lines in absorption, along with a broad He\,I $\lambda$4471 dip and possible hydrogen absorption lines. With the exception of the hydrogen features, it is very similar in appearance to the spectrum of the warm DQ SDSS$\,$J1036$+$6522.  
 \item OW$\,$J1753--3107 has a temperature of $T_{\rm eff} =$ \textcolor{black}{15430} K, and a carbon abundance of $\log{(N(C)/N(He))} =$ \textcolor{black}{--1.2}. As in the case of SDSS$\,$J1036$+$6522, these values are lower than those of known hot DQs. Thus, OW$\,$J1753--3107 can be considered to be a warm DQ, believed to exist in a transition state between the hot carbon-enriched DQ and cooler helium-dominant DQ.  
  \item  The current sample of known warm and hot DQs appear to reside in a distinct region of colour-colour space. However it is currently not possible to differentiate between hot DQs and warm DQs using their location in this plane.   
  \item  The mean magnetic field strength of OW$\,$J1753--3107 is B$_{z} =$2.1 MG, slightly lower than that of SDSS$\,$J1036$+$6522. The sinusoidal pulse shape of OW$\,$J1753--3107 provides further evidence to suggest that there is no relationship between non-linearities in the pulse shape of warm or hot DQs and their magnetic field strength.
  \item The origin of the 35-min variation seen in OW$\,$J1753--3107 is currently believed to be due to the rotation of the white dwarf. A pulsational model has been ruled out since there is no evidence for multi-mode variations, and the period is longer than expected from the pulsational periods observed in other white dwarfs. Furthermore, there is no evidence of the emission lines in its spectrum that would indicate it is an interacting binary star system undergoing accretion. The lack of significant radial velocity variations at the 35-min period, to an amplitude limit of $<$ 5 km/s, appears to rule out that OW J1753--3107 is an ultra-compact binary system.   
  \item Dedicated surveys are needed in order to increase and characterise the known population of hot and warm DQ. This will help to answer the most pressing questions in this research field, such as the origin of the variations and the type of evolutionary channels hot and warm DQ may follow. 
 \end{itemize}

\section*{Acknowledgements}
  The authors gratefully acknowledge funding from the Erasmus Mundus
 Programme SAPIENT, the National Research Foundation of South Africa
 (NRF), the Nederlandse Organisatie voor Wetenschappelijk Onderzoek
 (the Dutch Organisation for Science Research), Radboud University, and 
the University of Cape Town. P.D. acknowledge support from
NSERC (Canada). Armagh Observatory is core funded by the Northern Ireland Executive. The ESO observations used in this paper are based on observations made with ESO Telescopes at the La Silla Paranal Observatory under programme ID 093.D-0753(A) as part of the Dutch GTO time on OmegaCAM, and  177.D-3023 (VPHAS+). This paper uses observations made at the South African Astronomical Observatory (SAAO). Some of the observations reported in this paper were obtained with the Southern African Large Telescope (SALT) under the program 2016-1-SCI-015 (PI: Sally Macfarlane). This research is supported by the NWO/NRF Bilateral agreement supporting astronomical research. 
We thank the anonymous referee for the useful comments which have helped to improve the paper.



\bibliographystyle{mnras}
\bibliography{references}

\bsp	
\label{lastpage}
\end{document}